\documentstyle[twoside,12pt]{article}
\begin{document}
\pagestyle{empty}
%%%%%%%%%%%%%%%%%%%%%%%%%%%%%%%%%%%%%%%%
%%%%%%%%%%%%%%%\def\baselinestretch{1.5}
\catcode`@=11
\def\marginnote#1{}
%%%%%%%%%%%%%%%%%%%%%%%%%%%%%%%%%%%%%%%%%
%
%
\hyphenation{un-der-ground}
\hyphenation{tem-pe-ra-ture}
\hyphenation{tem-pe-ra-tures}
\hyphenation{dis-tri-bu-tions}
\hyphenation{dis-cri-mi-na-tor}

\begin{center}
{\Large \bf Quasiparticle and phonon propagation in Superheated Superconducting
 Granules after nuclear recoils.}
\end{center}
\vspace{.5cm}
A. Gabutti \\
{\it Laboratory for High Energy Physics, University of Bern,
Sidlerstrasse 5, CH 3012 Bern, Switzerland} \\
 \\
\begin{center}
\section*{Abstract}
\end{center}
\textwidth=5in
\baselineskip=12pt
%\fontsize{10}{12}
\small
\begin{quote}
The propagation of the excess of quasiparticles and phonons produced by a
nuclear recoil inside Sn and Zn superheated superconducting granules will be
discussed. The decay towards equilibrium of the initial disturbance is
assumed to be a thermal diffusion process described by a set of coupled
heat flow equations for the effective quasiparticle and phonon temperatures.
The solution is carried out analytically for a point source located
anywhere inside the superconducting granule with the initial energy
distributed in both quasiparticle and phonon systems.
The calculated time delay between the neutron interaction and the nucleation
of the phase transition will be compared to the time delay distributions
obtained by irradiating Zn and Sn SSG detectors with a 70Me$\!$V neutron
beam.\end{quote}

\textwidth=6in
\baselineskip=14pt
\normalsize
\section{Introduction}

Irradiation tests of SSG detectors  with a 70Me$\!$V neutron beam
have been used to study the delay in time between the neutron elastic
scattering and the phase transition of a single granule inside the
detector.  Typical delay times of $\sim$100ns and  $\sim$500ns have been
measured in Sn and Zn SSG respectively, as it is discussed
elsewhere in these proceedings$^{1}$.
% \cite{ref:klausi}

After a neutron elastic scattering, the deposited energy is
transferred to electrons and to the lattice by the recoiling nucleus.
The initial disturbance is localized at the
interaction point and produces an excess of quasiparticles and phonons in
a state very far from equilibrium where the kinetic energy of the
quasiparticles is larger than $kT_{c}$.
After a few $ps$ this hot electron
gas exhibits a local equilibrium with the phonons and a slower relaxation
process dominated by heat diffusion takes place$^{2}$.
% \cite{ref:gray}.

The heat propagation in superconducting spheres is
discussed in  \mbox{Ref. 3} for an ionizing particle
depositing energy only in the quasiparticle system as a primary excitation.
These results can not be applied to the case of nuclear recoils where a
large fraction of the energy is transferred to the phonons as a primary
excitation.
In the present work we will consider the heat diffusion after a nuclear
recoil interaction
located anywhere inside the superconducting granule and
with the initial energy distributed in both quasiparticle and phonon systems.

\section{Thermal Diffusion}
\label{chapter1}

The decay towards equilibrium of the initial disturbance is  described by
a set of coupled heat flow equations$^{3}$ for the effective quasiparticle
($T_{e}$) and phonon temperatures ($T_{p}$):
	\begin{eqnarray} \label{eq:diff}
& & \frac{\partial T_{e}}{\partial t} =  D_{e} \nabla^{2} T_{e} -\frac{1}
{\tau_{e}} (T_{e}-T_{p}) + A_{e} \delta_{e}(\vec{r_{o}},t_{o})\\
& &\frac{\partial T_{p}}{\partial t} =  D_{p} \nabla^{2} T_{p} -\frac{1}
{\tau_{p}} (T_{p}-T_{e}) -\frac{1}{\tau_{s}} (T_{p}-T_{b})+
A_{p} \delta_{p}(\vec{r_{o}},t_{o})\nonumber
	\end{eqnarray}
where $D_{e}$ and $D_{p}$ are the thermal diffusivities, $\tau_{e}$ and
$\tau_{p}$ are the quasiparticle and phonon lifetimes and $T_{b}$
the bath temperature. Since SSG detectors are made of
granules embedded in a dielectric material (plasticine or $Al_{2}O_{3}$
powder) the heat transfer to the bath is due only to phonons crossing the
granule surface with a rate 1/$\tau_{s}$.
The initial disturbance is assumed to be a point source in
$\vec{r_{o}}$ at the time $t_{o}$
described by the delta functions $\delta_{e}(\vec{r_{o}},t_{o})$ and
$\delta_{p}(\vec{r_{o}},t_{o})$ with $A_{e}$ and $A_{p}$ accounting for the
initial energy share between the two systems.

Equation \ref{eq:diff} can be solved analytically using Fourier transforms
and the eigenvalues method$^{4}$ with the thermal
diffusivities and the carrier lifetimes being constant in temperature.
Due to the spherical symmetry of the problem, the spatial eigenfunctions of
the solution are the $even$ and $odd$ spherical harmonic functions
$\Psi_{m,n,l}(\vec{r})$ with the indexes $n$ and $m$ referring to the
order of the spherical Bessel functions and Legendre polynomes
respectively. The boundary conditions on the granule surface are used to
evaluate the positive zeros of the first derivative of the Bessel functions
(index $l$).
The final expression for the quasiparticle temperature is:
	\begin{eqnarray} \label{eq:Te}
T_{e}(\vec{r},t)&=&T_{b} + F_{e}(t) \\
& & + V \sum_{m=0}^{\infty} \sum_{n=0}^{\infty}
\sum_{l=0}^{\infty} \big[ \Psi_{m,n,l}^{e}(\vec{r}) \Psi_{m,n,l}^{e}
(\vec{r_{o}}) + \Psi_{m,n,l}^{o}(\vec{r}) \Psi_{m,n,l}^{o}(\vec{r_{o}})
\big] F_{e}(l,t)\nonumber
        \end{eqnarray}
with $V$ the granule volume. A similar expression holds for $T_{p}(\vec{r},t)$
replacing $F_{e}$ with $F_{p}$.
The time dependent terms $F_{e}$ and $F_{p}$ are
linear combinations of the initial quasiparticle ($T_{e}^{i}$) and phonon
($T_{p}^{i}$) temperatures and exhibit a mixture of relaxation and diffusion
terms. The terms $F_{e}(t)$ and $F_{p}(t)$ are calculated from
$F_{e}(t,l)$ and $F_{p}(t,l)$ with $l\!=\!0$ and do not
depend on the spatial coordinates.
The derivation of \mbox{Eq. \ref{eq:Te}} is discussed in
\mbox{Ref. 5}.

At the beginning of the diffusion process the rise in
temperature is localized at the interaction point with
$T_{e}$ and $T_{p}$  proportional to the partition of the initial energy
between the two systems given by the conditions
$F_{e}(0,l)\!=\!F_{e}(0)\!=\!T_{e}^{i}$ and
$F_{p}(0,l)\!=\!F_{p}(0)\!=\!T_{p}^{i}$.
The terms $F_{e}(l,t)$ and $F_{p}(l,t)$ decay
exponentially to zero with time and at the end of the diffusion process
$(t\!=\! \infty)$
the equilibrium temperatures are
$T_{e}\!=\!T_{p}\!=\!T_{\infty}$ with $F_{e}(\infty)\!=\!F_{p}(\infty)$.
The energy conservation inside the granule, in the case of no heat
transfer to the bath $(1/\tau_{s}\!=\!0)$, impose the conditions:
	\begin{equation}\label{eq:Tinfi}
E_{r} = V \int_{T_{b}}^{T_{\infty}} (C_{es}+ C_{ph})\; dT \qquad ,  \qquad
\frac{\tau_{p}}{\tau_{e}} = \frac{T_{e}^{i} - T_{\infty}}
{T_{\infty} - T_{p}^{i}}
	\end{equation}
The initial quasiparticle and phonon temperatures $T_{e}^{i}$ and
$T_{p}^{i}$ are evaluated from the integral of the specific heat in the
superconducting state:
	\begin{equation}\label{eq:Tei}
E_{r} f = V \int_{T_{b}}^{T_{e}^{i}} C_{es} \;
dT \qquad ,  \qquad
E_{r} (1-f) = V \int_{T_{b}}^{T_{p}^{i}} C_{ph} \; dT
	\end{equation}
where $f$ is the fraction of the  recoil energy $E_{r}$ transferred to
electrons by neutron elastic scatterings and can be evaluated from
\mbox{Ref. 6}. Typical values of $f$ in Sn and Zn
absorbers are $\sim$0.22 and $\sim$0.4 for recoil energies of 5ke$\!$V and
100ke$\!$V respectively.

The lifetimes ratio and the initial temperatures defined in
\mbox{Eq. \ref{eq:Tinfi}}
and \mbox{Eq. \ref{eq:Tei}}, differ from the expression used in
previous works$^{3}$  where it was assumed that
$\tau_{p}/\tau_{e}\!=\!C_{ph}/C_{es}$ and $T_{e}^{i}\!=\!E_{r}/(VC_{es})$
with the specific heats constant in temperature. This approximation holds
only at high bath temperatures where the relative temperature rise is
small. At low temperatures where $C_{es}\!\ll\!C_{ph}$, the values
of $\tau_{p}$ evaluated from the ratio of the specific heats are an order of
magnitude bigger than $\tau_{e}$.
The carrier lifetime ratio defined in \mbox{Eq. \ref{eq:Tinfi}}
depends on the recoil energy and on the partition of the initial energy.

\section{Rise in temperature after a nuclear recoil}

In order to compare the calculations with the
neutron irradiation tests, we considered Sn and Zn granules with diameters
15$\mu$m and 30$\mu$m respectively. The quasiparticle diffusivity
was extrapolated from reference values$^{7}$  using the
normal state electrical resistivity measured$^{8}$ on granules
similar to the ones used in the irradiation tests.
The phonon thermal diffusivity is $D_{p}\!\ll\!D_{e}$ and can not be
extrapolated from measurements. We performed computations of the
quasiparticle and phonon temperatures either neglecting $D_{p}$ or
with $D_{p}\!=\!D_{e}/100$
without obtaining substantial differences in the thermal diffusion process.
The lifetimes $\tau_{e}$ were evaluated from$^{9}$
at the temperature $T_{\infty}$ for quasiparticle energies
two times the superconducting gap.
Typical values for $\tau_{e}$ are $\sim$300ns in Zn and $\sim$1ns
in Sn. The phonon lifetimes  $\tau_{p}$ were obtained from
\mbox{Eq. \ref{eq:Tinfi}}.
The rate $1/\tau_{s}$ at which  phonons cross the granule surface was
neglected since $\tau_{s}$ is of the order of few $\mu$s.
Nuclear recoils inside the superconducting granule were simulated
considering a three dimensional grid of 72 equally spaced interaction points
located in the half sphere with the coordinate $\phi$ in the interval
$0 \!- \! \pi$.
For each interaction point, the quasiparticle and phonon temperatures
were calculated for a  point on the granule surface
located on the equatorial plane at $\phi\!=\!\pi /2$.

The calculated temperatures are plotted in Fig. 1
for a recoil energy of 5ke$\!$V deposited in a Zn granule at the bath
temperature of 50mK.
Due to the differences between the lifetimes in Zn,
$\tau_{e}$=300ns and $\tau_{p}$=50ns, the initial perturbation is
transferred to the quasiparticles system in a shorter time scale.
At the beginning of the diffusion
process, the temperature rise depends on the location of the interaction
point inside the granule and temperatures higher than the final value
$T_{\infty}$ can be reached for interaction points close to the surface.
At the end of the diffusion, all the temperatures converge to $T_{\infty}$.
In Sn granules, the time scale for the relaxation process is shorter
because $\tau_{e}\!\sim\!\tau_{p}\!\sim\!1$ns and the quasiparticle and phonon
temperatures exhibit the same time dependence.

\section{Time delay}

The nuclear recoil interactions were simulated considering a three
dimensional grid of 122 equally spaced points located in the whole granule
volume. For each interaction point
the temperatures were computed with time steps of 4ns from 0 up to
400ns in Sn and $1.5\mu$s in Zn.

The energy threshold of a single granule is related to the magnetic
threshold $h\!=\!1\!-\!B_{a}/B_{sh}(T_{b})$ where $B_{sh}$ is the granule
superheating field and $B_{a}$ is the strength of the applied magnetic
field. From the phase diagram it is possible to relate the magnetic threshold
$h$ to the temperature $T_{c}^{\ast}$ needed to fulfill the condition
$B_{a}\!=\!B_{sh}(T_{c}^{\ast})$.

Previous irradiation experiments on single granules with radioactive
sources$^{10}$ have shown that
the  phase transition tends to nucleate in a small portion of the granule
surface (nucleation center) and spread afterwards into the full granule
volume. This effect was found to be less evident in Zn than in Sn granules.
Measurements have also shown that there is a dependence of the
the phase transition field on the crystallographic orientation of the
granule with respect to the applied magnetic field$^{11}$.
Since SSG are made of a collection of granules, it is quite
difficult to define a criteria for the occurance of the phase
transition valid for all the granules inside the detector.

To evaluate the time elapsing from the interaction to the nucleation of
the phase transition, we monitored the quasiparticle temperature on 5
equidistant points on the granule equator and recorded the time for which
the condition $T_{e}\!\ge\!T_{c}^{\ast}$ was simultaneously satisfied on the 5
points.
This is a simplified assumption, because the nucleation of the
phase transition is assumed to be along the full equatorial plane neglecting
the possibility of having a more localized nucleation center or a combination
of the two mechanisms.
The measured SSG superheating field distributions$^{12}$ were
used to evaluate the  probability distribution of the transition
temperatures $T_{c}^{\ast}$.

To compare with the measurements, the calculated time distributions were
weighted with the neutron elastic scattering cross section and with
the detection efficiency of the neutron counter used in the
experiments$^{12}$.
Despite the simplified assumption used to define the occurance of the
phase transition, the calculation well reproduce the time scale of the
experimental results as shown in Fig. 2.
It is important to note that the measured distribution have a
constant time offset due to the electronic delays$^{1}$.

In the case of Zn granules, where the diffusion time is longer, there is
a difference in the shape of the calculated distributions  for recoil
energies in the ranges 5-15ke$\!$V and 40-60ke$\!$V.
At low recoil energies
the theoretical distribution is shifted toward longer times in
agreement with the measurements.
In Sn, the quasiparticle relaxation times are shorter and both the
calculated and measured time delay distributions are within 200ns.

\section{Conclusions}
\label{chapter4}

An analytical expression to describe the heat propagation inside
superconducting granules after nuclear recoils was derived.
The calculations differ from previous works because the initial energy
is shared in both quasiparticle and phonon systems.
The elapsed time from the nuclear recoil interaction to the nucleation
of the phase transition was calculated for Sn an Zn granules.
The calculations are in good agreement with the distributions measured  in
irradiation tests of SSG with a 70Me$\!$V neutron beam.

\section*{Acknowledgments}
I would like to thank K. Pretzl and K. Schmiemann from the University of
Bern for helpful discussions.
This work was supported by the Schweizerischer Nationalfonds zur
Foerderung der wissenschaftlichen Forschung and by the Bernische Stiftung
zur Foerderung  der wissenschaftlichen Forschung an der Universitaet Bern.

%% polloref:
\section*{References}
\begin{enumerate}
\item M. Abplanalp et {\it al.} in these proceedings. \vspace{-3mm}
\item K. Gray in {\it Superconductive particles detectors}, ed.
A. Barone (World Scientific, 1987) p.1. \vspace{-3mm}
\item N. Perrin, J. Physique {\bf47} (1986) 1939. \vspace{-3mm}
\item P.M. Morse and H. Feshbach, {\it Methods of theoretical
physics}, (McGraw-Hill Book Company, New York, 1953) p. 1468. \vspace{-3mm}
\item A. Gabutti to be published \vspace{-3mm}
\item J. Lindhard et {\it al.}, Mat. Fys. Medd. Dan. Vid. Selsk.
{\bf33/10} (1963) 1. \vspace{-3mm}
\item Y.S. Touloukian et {\it al.}, {\it Thermal conductivity},
(IF/Plenum, New York, 1970).\vspace{-3mm}
\item M. Furlan et {\it al.}, Nucl. Inst. and Meth. {\bf A338} (1994) 544.
\vspace{-3mm}
\item S.B. Kaplan et {\it al.}, Phys. Rev. {\bf B14/11} (1976) 4854.
 \vspace{-3mm}
\item M. Frank et {\it al.}, Nucl. Inst. and Meth. {\bf A287} (1990) 583.
\vspace{-3mm}
\item K. Pretzl, Particle World {\bf 1/6} (1990) 153. \vspace{-3mm}
\item M. Abplanalp et {\it al.},  to be published. \vspace{-3mm}
\end{enumerate}
\section*{Figure Captions}
\begin{itemize}
\item [Fig. 1] Quasiparticle (q.p.) and phonon temperatures versus time for
5ke$\!$V nuclear recoils in Zn. The interactions are
distributed in a three dimensional grid of 72 equally spaced points
located inside the half sphere $\phi\!=\! 0\!-\!\pi$. The
temperatures are monitored on the granule equator at the point
$\phi\!=\!\pi /2$.
\item [Fig. 2] Calculated (upper) and measured$^{1}$ (lower)
distributions of the time delay between nuclear recoils and phase
transitions in Zn and Sn granules with magnetic thresholds $h$=2$\%$ and
$h$=1$\%$ respectively. The measured
distributions have a constant time offset due to the electronic readout.
\end{itemize}
\end{document}